\documentclass[10pt,letterpaper,conference]{ieeeconf}

\usepackage{latexsym, amssymb, amsmath}
\usepackage[dvips]{graphicx} 
\usepackage{enumerate}
\usepackage{flushend} 
\usepackage{mathrsfs} 
\usepackage{stfloats}
\usepackage{cite}

\allowdisplaybreaks[3] 

\newtheorem{prop}{Proposition} 
\newtheorem{proposition}[prop]{Proposition} 
\newtheorem{lem}{Lemma} 
\newtheorem{lemma}[lem]{Lemma} 
\newtheorem{thm}{Theorem} 
\newtheorem{theorem}[thm]{Theorem} 
\newtheorem{cor}{Corollary} 
\newtheorem{corollary}[cor]{Corollary} 
\newtheorem{defn}{Definition} 
\newtheorem{definition}[defn]{Definition} 
\newtheorem{exmp}{Example} 
\newtheorem{example}[exmp]{Example} 
\newtheorem{exam*}{Example}

\def\custombibliography#1{
 \normalsize
\section*{\centering References}
 \list
 {[\arabic{enumi}]}{\settowidth\labelwidth{[#1]}\leftmargin\labelwidth
 \setlength{\itemsep}{.1em}
 \advance\leftmargin\labelsep
 \usecounter{enumi}}
 \def\newblock{\hskip .11em plus .33em minus -.07em}
 \sloppy
 \sfcode`\.=1000\relax}

\def\L2{{\cal L}_2}
\def\begar{\begin{array}}
\def\endar{\end{array}}
\def\begce{\begin{center}}
\def\endce{\end{center}}
\def\begco{\begin{cor}}
\def\endco{\end{cor}}
\def\begde{\begin{defn}}
\def\endde{\end{defn}}
\def\begdes{\begin{description}}
\def\enddes{\end{description}}
\def\begdi{\begin{displaymath}}
\def\enddi{\end{displaymath}}
\def\begdis{\begin{eqnarray*}}
\def\enddis{\end{eqnarray*}}
\def\begen{\begin{enumerate}}
\def\enden{\end{enumerate}}
\def\begeq{\begin{equation}}
\def\endeq{\end{equation}}
\def\begeqa{\begin{eqnarray}}
\def\endeqa{\end{eqnarray}}
\def\begex{\begin{exmp}}
\def\endex{\end{exmp}}
\def\begfig{\begin{fig}}
\def\endfig{\end{fig}}
\def\begit{\begin{itemize}}
\def\endit{\end{itemize}}
\def\begle{\begin{lem}}
\def\endle{\end{lem}}
\def\begpro{\begin{prop}} 
\def\endpro{\end{prop}} 
\def\begth{\begin{thm}}
\def\endth{\end{thm}}

\def\begres{\noindent{\bf Remarks}:\begin{enumerate}}
\def\endres{\end{enumerate} \par}
\def\begpr{\begin{proof}} 
\def\endpr{\end{proof}}

\newcommand\bull{\vrule height .9ex width .8ex depth -.1ex } 

\newcommand\re{\rm I\! R}
\newcommand\cdcout[1]{} 

\DeclareMathOperator{\diag}{diag}  

\newcommand{\rv}[1]{\boldsymbol{#1}} 
\newcommand{\RomanNumber}[1]{\uppercase\expandafter{\romannumeral #1}}
\newcommand{\romannumber}[1]{\lowercase\expandafter{\romannumeral #1}}

\DeclareMathAlphabet{\mathpzc}{OT1}{pzc}{m}{it}

\def\1{\rv 1} 

\IEEEoverridecommandlockouts                              
\title{Additive Networks of Chen-Fliess Series: \\ Local Convergence and Relative Degree}

\author{W.~Steven Gray$^{\dag}$\thanks{$^{\dag}$Department of Electrical and Computer
Engineering, Old Dominion University, Norfolk, Virginia 23529, USA, email: sgray@odu.edu.}
$\;\;\;$
Luis A.~Duffaut Espinosa$^{\ddag}$\thanks{$^{\ddag}$Department of Electrical and
Biomedical Engineering, University of Vermont, Burlington, Vermont 05405 USA, email: lduffaut@uvm.edu.}
$\;\;\;$
Kurusch Ebrahimi-Fard$^{\S}$\thanks{$^{\S}$Department of Mathematical Sciences,
		Norwegian University of Science and Technology (NTNU),
		7491 Trondheim, Norway, email: kurusch.ebrahimi-fard@ntnu.no.}
}

\usepackage{cite} 
\usepackage{color} 
\usepackage{graphicx} 
\usepackage{latexsym,amssymb,amsmath,amsfonts} 
\usepackage{mathrsfs} 
\usepackage{multicol}
\usepackage{fixltx2e} 
\usepackage{multirow} 
\usepackage{longtable} 
\usepackage{booktabs}

\renewcommand{\baselinestretch}{1}
\allowdisplaybreaks[4]
\definecolor{Light}{gray}{0.85}

\def\abs#1{\left\vert #1 \right\vert}
\def\allpolyell{\mbox{$\re^{\ell}\langle X \rangle$}}
\def\allpolyx0degn{\mbox{$P_n$}}

\def\allseries{\mbox{$\re\langle\langle X \rangle\rangle$}}

\def\allseriesell{\mbox{$\re^{\ell} \langle\langle X \rangle\rangle$}}

\def\allseriesmLC{\mbox{$\re^{m}_{LC}\langle\langle X \rangle\rangle$}}

\def\allseriesellLC{\mbox{$\re^{\ell}_{LC}\langle\langle X \rangle\rangle$}}

\def\bull{\rule{0.08in}{0.08in}} 

\newcommand{\comment}[1]{} 

\def\diag{{\rm diag}}

\def\Endallseries{{\rm End}(\allseries)}

\def\eqref#1{(\ref{#1})} 

\def\lieseries{\widehat{\mathcal L}(X)}

\def\mbf#1{\hbox{\mathversion{bold}$#1$}} 

\def\norm#1{\left\Vert#1\right\Vert}
\def\notin{{\not\in}}

\def\openbull{\framebox[0.08in][c]{$\;$}} 

\def\re{{\mathbb R}} 

\def\shuffle{{\scriptscriptstyle \;\sqcup \hspace*{-0.05cm}\sqcup\;}}

\def\supp{{\rm supp}}

\def\begals{\[\begin{aligned}}
\def\endals{\end{aligned}\]}
\def\begce{\begin{center}}
\def\endce{\end{center}}
\def\begar{\begin{array}}
\def\endar{\end{array}}
\def\begeq{\begin{equation}}
\def\endeq{\end{equation}}
\def\begdi{\begin{displaymath}}
\def\enddi{\end{displaymath}}
\def\begdis{\begin{eqnarray*}}
\def\enddis{\end{eqnarray*}}
\def\begeqa{\begin{eqnarray}}
\def\endeqa{\end{eqnarray}}
\def\begdes{\begin{description}}
\def\enddes{\end{description}}
\def\begit{\begin{itemize}}
\def\endit{\end{itemize}}
\def\begen{\begin{enumerate}}
\def\enden{\end{enumerate}}
\def\beglar{\left[\begin{array}}
\def\endrar{\end{array}\right]}
\def\begle{\begin{mylemma}}
\def\endle{\end{mylemma}}
\def\begde{\begin{mydefinition}}
\def\endde{\end{mydefinition}}
\def\begth{\begin{mytheorem}}
\def\endth{\end{mytheorem}}
\def\begco{\begin{mycorollary}}
\def\endco{\end{mycorollary}}
\def\begprop{\begin{myproposition}}
\def\endprop{\end{myproposition}}
\def\begex{\begin{myexample}}
\def\endex{\hfill\openbull \end{myexample} \vspace*{0.15in}}
\def\begexer{\begin{myexercise}}
\def\endexer{\end{myexercise}}

\def\begres{\noindent{\bf Remarks}:\begin{enumerate}}
\def\endres{\end{enumerate} \par}
\def\begpr{\noindent{\em Proof:}$\;\;$}
\def\endpr{\hfill\bull \vspace*{0.15in}}
\def\begtab{\begin{tabular}}
\def\endtab{\end{tabular}}
\def\rref#1{(\ref{#1})}

\def\allseriesA{\mbox{$\re\!\ll\!A\!\gg$}}
\def\allseriesA'{\mbox{$\re\!\ll\!A'\!\gg$}}

\def\shuff#1#2{\mathbin{
      \hbox{\vbox{\hbox{\vrule \hskip#2 \vrule height#1 width 0pt}\hrule}\vbox{\hbox{\vrule \hskip#2 \vrule height#1 width 0pt\vrule }\hrule}}}}
\def\shuffl{{\mathchoice{\shuff{5pt}{3.5pt}}{\shuff{5pt}{3.5pt}}{\shuff{3pt}{2.6pt}}{\shuff{3pt}{2.6pt}}}}
\def\shuffle{{\, \shuffl \,}}
\def\allseriesLCtensorn{\mbox{$\re_{LC}^{\otimes n}\langle\langle X \rangle\rangle$}}
\def\allseriesLCtensorm{\mbox{$\re_{LC}^{\otimes m}\langle\langle X \rangle\rangle$}}

\def\begce{\begin{center}}
\def\endce{\end{center}}
\def\begar{\begin{array}}
\def\endar{\end{array}}
\def\begeq{\begin{equation}}
\def\endeq{\end{equation}}
\def\begdi{\begin{displaymath}}
\def\enddi{\end{displaymath}}
\def\begdis{\begin{eqnarray*}}
\def\enddis{\end{eqnarray*}}
\def\begeqa{\begin{eqnarray}}
\def\endeqa{\end{eqnarray}}
\def\begdes{\begin{description}}
\def\enddes{\end{description}}
\def\begit{\begin{itemize}}
\def\endit{\end{itemize}}
\def\begen{\begin{enumerate}}
\def\enden{\end{enumerate}}
\def\beglar{\left[\begin{array}}
\def\endrar{\end{array}\right]}
\def\begle{\begin{lemma}}
\def\endle{\end{lemma}}
\def\begde{\begin{definition}}
\def\endde{\end{definition}}
\def\begth{\begin{theorem}}
\def\endth{\end{theorem}}
\def\begco{\begin{corollary}}
\def\endco{\end{corollary}}
\def\begprop{\begin{proposition}}
\def\endprop{\end{proposition}}
\def\begex{\begin{example}}
\def\endex{\hfill\openbull \end{example} \vspace*{0.1in}}
\def\begexer{\begin{exercise}}
\def\endexer{\end{exercise}}

\def\begres{\noindent{\bf Remarks}:\begin{enumerate}}
\def\endres{\end{enumerate} \par}
\def\begpr{\noindent{\em Proof:}$\;\;$}
\def\endpr{\hfill\bull \vspace*{0.05in}}
\def\begtab{\begin{tabular}}
\def\endtab{\end{tabular}}
\def\rref#1{(\ref{#1})}

\begin{document}

\maketitle

\begin{abstract}
Given an additive network of input-output systems where each
node of the network is modeled by a locally convergent Chen-Fliess series,
two basic properties of the network are established. First, it is
shown that every input-output map between a given pair of nodes has
a locally convergent Chen-Fliess series representation. Second, sufficient
conditions are given under which the input-output map between a pair
of nodes has a well defined relative degree as defined by
its generating series. This analysis leads to the conclusion
that this relative degree property is generic in a certain sense.
\end{abstract}

\section{Introduction}

Networks of nonlinear dynamical systems appear in many fields, especially in the natural sciences
where the nonlinearity is often a key feature in generating the observed behavior \cite{Golubitsky-Stewart_06,Jiang-Lai_19}.
The vast majority of analysis of such networks is done in a finite dimensional state space setting using coupled
systems of ordinary differential equations. In \cite{Gray-Ebrahimi-Fard_SCL21}, however, the authors describe an alternative
approach which uses only input-output models at each node of the network in the form of a locally convergent Chen-Fliess series \cite{Fliess_81,Fliess_83}.
These weighted infinite sums of iterated integrals provide a convenient algebraic framework for describing the
network's behavior without relying on any particular choice of coordinates as in the state space setting.
Series coefficients for each node can be estimated via system identification techniques \cite{Gray-etal_Auto20}.
Computational tools were developed in \cite{Gray-Ebrahimi-Fard_SCL21} to determine,
for example, how an input injected at one node affects the output observed at another node.
Nevertheless, there are still a number of
open questions regarding the basic properties of such networks. The focus here will be on so called {\em additive} networks, where the outputs
of the nodes are simply added together and injected into other nodes, including self-loops. Other classes of aggregation functions
such the multiplication of node outputs will not be addressed here.

This paper has two goals. The first goal to address the open problem stated in \cite{Gray-Ebrahimi-Fard_SCL21} regarding whether an additive network of
locally convergent Chen-Fliess series always yields mappings between nodes which have locally convergent Chen-Fliess series representations.
This hypothesis will be proved to be true and is independent of the network's topology.
The approach taken is to identify for a given network an associated {\em maximal network} whose
growth bounds on the coefficients of the generating series between nodes upper bound all the growth bounds of the original network and are much
easier to determine using conventional methods as presented in \cite{Sussmann_83}. The particular growth bound derived turns out to be exactly
equivalent to one discovered for a class of unity feedback systems described in \cite{Thitsa-Gray_12}.
The second goal
is to provide sufficient conditions under which the input-output map between a pair of nodes has well defined relative degree as defined by
its generating series \cite{Gray-etal_AUTO14,Gray-Venkatesh_SCL19}. A simple counterexample will be given first to show that
this property can fail to hold in certain situations.
The proofs of the sufficient conditions rely on identifying certain properties first described in \cite{Gray-Venkatesh_SCL19}
in relation to a subgraph connecting a given input node and output node.
It is also shown, however, that this relative degree property is {\em generic} in a certain sense.
Namely, if the generating series for every node has relative degree and the connection strengths between the nodes are random, then every node pair has
a generating series with well defined relative degree with probability one.
An obvious application for
this result is in the context of feedback linearization for networks \cite{Menara-etal_CDC20}, however, that application will not
be pursued here.

The paper is organized as follows. To keep the presentation as self-contained as possible, the required preliminaries are briefly
summarized in Section~\ref{sec:preliminaries}. The question regarding the convergence of Chen-Fliess series for
mappings between nodes is addressed in Section~\ref{sec:local-convergence}.
The subsequent section treats the property of relative degree. The paper's conclusions are summarized in the final section.

\section{Preliminaries}
\label{sec:preliminaries}

An {\em alphabet} $X=\{ x_0,x_1,$ $\ldots,x_m\}$ is any nonempty and finite set
of noncommuting symbols referred to as {\em
letters}. A {\em word} $\eta=x_{i_1}\cdots x_{i_k}$ is a finite sequence of letters from $X$.
The number of letters in a word $\eta$, written as $\abs{\eta}$, is called its {\em length}.
The empty word, $\emptyset$, is taken to have length zero.
The collection of all words having length $k$ is denoted by
$X^k$. Define $X^\ast=\bigcup_{k\geq 0} X^k$,
which is a monoid under the concatenation (Cauchy) product.
Any mapping $c:X^\ast\rightarrow
\re^\ell$ is called a {\em formal power series}.
Often $c$ is
written as the formal sum $c=\sum_{\eta\in X^\ast}\langle c,\eta\rangle\eta$,
where the {\em coefficient} $\langle c,\eta\rangle\in\re^\ell$ is the image of
$\eta\in X^\ast$ under $c$.
The {\em support} of $c$, $\supp(c)$, is the set of all words having nonzero coefficients.
A series $c$ is {\em proper} when $\emptyset\not\in\supp(c)$.
The set of all noncommutative formal power series over the alphabet $X$ is
denoted by $\allseriesell$. The subset of series with finite support, i.e., polynomials,
is represented by $\allpolyell$.
For any $c,d\in\allseries$, the scalar product is
$\langle c,d\rangle:=\sum_{\eta\in X^\ast} \langle c,\eta\rangle\langle d,\eta\rangle$, provided the sum is finite.
The set $\allseriesell$ is an associative $\re$-algebra under the concatenation product and an associative and commutative $\re$-algebra under
the {\em shuffle product}, that is, the bilinear product uniquely specified by the shuffle product of two words $x_i\eta,x_j\xi\in X^\ast$:
\begdi
	(x_i\eta)\shuffle(x_j\xi)=x_i(\eta\shuffle(x_j\xi))+x_j((x_i\eta)\shuffle \xi),
\enddi
where $x_i,x_j\in X$ and with $\eta\shuffle\emptyset=\emptyset\shuffle\eta=\eta$ \cite{Fliess_81}.

\subsection{Chen-Fliess series}

Given any $c\in\allseriesell$ one can associate a causal
$m$-input, $\ell$-output operator, $F_c$, in the following manner.
Let $\mathfrak{p}\ge 1$ and $t_0 < t_1$ be given. For a Lebesgue measurable
function $u: [t_0,t_1] \rightarrow\re^m$, define
$\norm{u}_{\mathfrak{p}}=\max\{\norm{u_i}_{\mathfrak{p}}: \ 1\le
i\le m\}$, where $\norm{u_i}_{\mathfrak{p}}$ is the usual
$L_{\mathfrak{p}}$-norm for a measurable real-valued function,
$u_i$, defined on $[t_0,t_1]$.  Let $L^m_{\mathfrak{p}}[t_0,t_1]$
denote the set of all measurable functions defined on $[t_0,t_1]$
having a finite $\norm{\cdot}_{\mathfrak{p}}$ norm and
$B_{\mathfrak{p}}^m(R)[t_0,t_1]:=\{u\in
L_{\mathfrak{p}}^m[t_0,t_1]:\norm{u}_{\mathfrak{p}}\leq R\}$.
Assume $C[t_0,t_1]$
is the subset of continuous functions in $L_{1}^m[t_0,t_1]$. Define
inductively for each word $\eta=x_i\bar{\eta}\in X^{\ast}$ the map $E_\eta:
L_1^m[t_0, t_1]\rightarrow C[t_0, t_1]$ by setting
$E_\emptyset[u]=1$ and letting
\[E_{x_i\bar{\eta}}[u](t,t_0) =
\int_{t_0}^tu_{i}(\tau)E_{\bar{\eta}}[u](\tau,t_0)\,d\tau, \] where
$x_i\in X$, $\bar{\eta}\in X^{\ast}$, and $u_0=1$. The
{\em Chen--Fliess series} corresponding to $c\in\allseriesell$ is
\begdi
y(t)=F_c[u](t) =
\sum_{\eta\in X^{\ast}} \langle c,\eta\rangle \,E_\eta[u](t,t_0) 
\enddi
\hspace*{-0.06in}\cite{Fliess_81}.
If there exist real numbers $K,M>0$ such that
\begdi
\abs{\langle c,\eta\rangle}\le K M^{|\eta|}|\eta|!,\;\; \forall\eta\in X^{\ast},
\enddi
then $F_c$ constitutes a well defined mapping from
$B_{\mathfrak p}^m(R)[t_0,$ $t_0+T]$ into $B_{\mathfrak
q}^{\ell}(S)[t_0, \, t_0+T]$ for sufficiently small $R,T >0$ and some $S>0$,
where the numbers $\mathfrak{p},\mathfrak{q}\in[1,\infty]$ are
conjugate exponents, i.e., $1/\mathfrak{p}+1/\mathfrak{q}=1$  \cite{Gray-Wang_02}.
(Here, $\abs{z}:=\max_i \abs{z_i}$ when $z\in\re^\ell$.) The set of all such
{\em locally convergent} series is denoted by $\allseriesellLC$, and $F_c$ is
referred to as a {\em Fliess operator}.

\subsection{System interconnections}

Given Fliess operators $F_c$ and $F_d$, where $c,d\in\allseriesellLC$,
the parallel and product connections satisfy $F_c+F_d=F_{c+d}$ and $F_cF_d=F_{c\shuffle d}$,
respectively \cite{Fliess_81}.
When Fliess operators $F_c$ and $F_d$ with $c\in\allseriesellLC$ and
$d\in\allseriesmLC$ are interconnected in a cascade fashion, the composite
system $F_c\circ F_d$ has the
Fliess operator representation $F_{c\circ d}$, where
the {\em composition product} of $c$ and $d$
is given by
\begdi
c\circ d=\sum_{\eta\in X^\ast} \langle c,\eta\rangle\,\psi_d(\eta)(\mathbf{1})
\enddi%
\hspace*{-0.07in}\cite{Ferfera_80}. Here $\mbf{1}$ denotes the monomial $1\emptyset$, and
$\psi_d$ is the continuous (in the ultrametric sense) algebra homomorphism
from $\allseries$ to the vector space endomorphisms on $\allseries$, $\Endallseries$, uniquely specified by
$\psi_d(x_i\eta)=\psi_d(x_i)\circ \psi_d(\eta)$ with
$ 
\psi_d(x_i)(e)=x_0(d_i\shuffle e),
$
$i=0,1,\ldots,m$
for any $e\in\allseries$,
and where $d_i$ is the $i$-th component series of $d$
($d_0:=\mbf{1}$). By definition,
$\psi_d(\emptyset)$ is the identity map on $\allseries$.

\subsection{Relative degree of a generating series}

Let $X=\{x_0,x_1\}$.
Following \cite{Gray-etal_AUTO14}, a series $c\in\allseries$ has relative degree $r$ if and only if it
has the decomposition
\begdi 
c=c_N+Kx_0^{r-1}x_1+x_0^{r-1}e
\enddi
for some $K\neq 0$ and proper $e\in\allseries$ with $x_1\not\in\supp(e)$.
This definition of relative degree is consistent with the classical definition
whenever $y=F_c[u]$
is realizable \cite{Gray-etal_AUTO14,Gray-Ebrahimi-Fard_SIAM17}.
The following results
will be of central
importance in the work that follows.

\begth \cite{Gray-Venkatesh_SCL19} \label{th:r-parallel-sum-connection}
If $c,d\in\allseries$ have distinct relative degrees $r_c$ and $r_d$, respectively, then $c+d$ has relative
degree $\min(r_c,r_d)$. On the other hand, if $r_c = r_d =: r$, then $c+d$ has relative degree $r$ if and
only if  $\langle c,x_{0}^{r-1}x_{1}\rangle + \langle d,x_{0}^{r-1}x_{1}\rangle \neq 0$.
\endth

\begco \label{co:relative-degree-multi-sum-distinct}
If $c_1,c_2,\ldots,c_m$ have relative degree $r_1,r_2,\ldots,r_m$, respectively,
with $r_i\neq r_j$ when $i\neq j$,
then the relative degree of $c_1+c_2+\cdots+c_m$ is $\min_i(r_i)$.
\endco

\begco \label{co:relative-degree-multi-sum-not-distinct}
Suppose $c_1,c_2,\ldots,c_m$ have relative degree $r_1,r_2,\ldots,r_m$, respectively.
Let $s_j$ denote the multiplicity of relative degree $r_j$.
If for each $s_j>1$ the series $c_{k_1},c_{k_2},\ldots,c_{k_{s_j}}$ having relative degree $r_j$
satisfy
\begdi
\langle c_{k_1},x_0^{r_j-1}x_1\rangle+\langle c_{k_2},x_0^{r_j-1}x_1\rangle+\dots+\langle c_{k_{s_j}},x_0^{r_j-1}x_1\rangle\neq 0,
\enddi
then the relative degree of $c_1+c_2+\cdots+c_m$ is $\min_i(r_i)$.
\endco

\begth \cite{Gray-Venkatesh_SCL19} \label{th:relative-degree-casecade}
If $c,d\in\allseries$ have relative degrees $r_c$ and $r_d$, respectively,
then $r_{c\circ d}$ has relative degree $r_c+r_d$.
\endth

\subsection{Formal realizations and representations}

It is shown in \cite{Kawski-Sussmann_97} that a given Chen-Fliess series $y=F_c[u]$ can be written in terms
of a state $z$ evolving on a formal Lie group $\mathcal{G}(X)$ with Lie algebra $\lieseries$ and output
map $y=\langle c,z\rangle$. This notion of a {\em universal control system} was generalized in
\cite{Gray-Ebrahimi-Fard_SCL21} as follows to describe networks of Chen-Fliess series.

\begde
Let $V_i$ be a vector field on $\mathcal{G}^n(X):=\mathcal{G}(X)\times\mathcal{G}(X)\times\cdots\times \mathcal{G}(X)$, $i=0,1,\ldots,m$ with
\begin{align*}
V_i&:{\mathcal G}^n(X)\rightarrow T_z{\mathcal G}^n(X) \\
&z=(z_1,\ldots,z_n)\mapsto V_i(z)=(V_{i1}(z)z_1,\ldots,V_{in}(z)z_n),
\end{align*}
where $V_{ij}(z(t))\in \widehat{\mathcal L}(X)$.
The $j$-th component of the corresponding state equation on ${\mathcal G}^n(X)$ is
\begdi 
\dot{z}_j=\sum_{i=0}^m V_{ij}(z)z_j u_{ij} , \;\; z_j(0)=z_{j0}.
\enddi
Given $\hat{c}_{k}\in\allseriesLCtensorn$, $k=1,2,\ldots,\ell$, the $k$-th output equation is defined to be
\begdi 
y_{k}=\hat{c}_{k}(z).
\enddi
Collectively, $(V,z_0,\hat{c})$ is a {\em formal realization} on ${\mathcal G}^n(X)$ of the formal input-output map
$u\mapsto y$.
\endde

Analogous to the standard finite dimensional theory \cite{Isidori_95,Nijmeijer-vanderSchaft_90}, a series $c\in\allseriesell$ is said to have a {\em formal representation} when
there exists a formal realization with the property that every coefficient of $c$ can be written in terms of iterated Lie derivatives of the vectors
fields acting on the output map and evaluated at $z_0$, i.e., $\langle c,x_{i_1}\cdots x_{i_k}\rangle=L_{V_{i_k}}\cdots L_{V_{i_1}}\hat{c}(z_0)$.

\section{Additive Networks of Chen-Fliess Series: Local Convergence}
\label{sec:local-convergence}

In this section it is shown that every network of additively interconnected locally convergent Fliess operators has the
property that the input-output maps between any two nodes can be represented by a locally convergent Fliess operator.
The first definition describes the specific class of networks under consideration.

\begde
A set of $m$ single-input, single-output Chen-Fliess series mapping $u_i$ to $y_i$ with generating series
$c_i\in\re_{LC}\langle\langle X_i\rangle\rangle$, where $X_i=\{x_0,x_i\}$
is said to be an {\em additively interconnected network} ${\mathcal N}_m$
with weighting matrix $W\in\re^{m\times m}$
if $u_i=v_i+\sum_{j=1}^mW_{ij}y_j$, $i=1,2,\ldots,m$.
\endde

A network ${\mathcal N}_m$ can therefore be viewed as a directed graph connecting $m$ nodes,
where the $i$-th node corresponds to a Chen-Fliess series with generating series $c_i$, $i=1,2,\ldots,m$.
Henceforth, it will be assumed that the connection weights are normalized so that $W_{ij}\in[0,1]$, $i,j=1,2,\ldots,m$.
The following theorem follows directly from Theorem 5.1 in \cite{Gray-Ebrahimi-Fard_SCL21}.

\begth \label{th:additive-interconnections}
The input-output map $v_i\mapsto y_j$ in any additively interconnected network ${\mathcal N}_m$
has generating series $d_{ji}\in\re\langle\langle X_i\rangle\rangle$ which can be computed from a formal
representation in terms of the vector fields
\begin{align*}
V_0(z)&=
\left[
\begin{tabular}{p{0.5cm}}
\hspace*{-0.1cm}$x_0z_1$ \\
\hspace*{-0.1cm}$x_0z_2$ \\
\hspace*{0.2cm}\vdots \\
\hspace*{-0.1cm}$x_0z_m$
\end{tabular}
\right]+\diag(x_1z_1,\ldots,x_m z_m)W
\left[
\begin{tabular}{p{1cm}}
\hspace*{-0.02cm}$\langle c_1,z_1\rangle$ \\
\hspace*{-0.02cm}$\langle c_2,z_2\rangle$ \\
\hspace*{0.4cm}\vdots \\
\hspace*{-0.13cm}$\langle c_m,z_m\rangle$
\end{tabular}
\right] \\
V_i(z)&=x_iz_i\mbf{e}_i
\end{align*}
acting on
$\hat{c}_j=\mbf{1}\otimes\cdots\otimes\mbf{1}\otimes c_j\otimes\mbf{1}\cdots\otimes\mbf{1}\in\allseriesLCtensorm$ ($c_j$ appears in the $j$-th position)
and evaluated at $z_{j0}=\mbf{1}$, $i,j=1,2\ldots,m$.
\endth

The next theorem states the main convergence result concerning additive networks.

\begth
If ${\mathcal N}_m$ is an additively interconnected network where the generating series for each node $c_i\in\re_{LC}\langle\langle X_i\rangle\rangle$,
then the generating series for every input-output map $d_{ji}\in\re_{LC}\langle\langle X_i\rangle\rangle$. More specifically, if $K_i,M_i$ denote the growth
constants for $c_i$, then for all $i,j=1,2,\ldots,m$
\begdi
\abs{\langle d_{ji},\eta\rangle}<KM^{\abs{\eta}}\abs{\eta}!,\;\;\forall \eta\in X^\ast
\enddi
for some $K>0$ and any $M>M_{\rm inf}$, where
\begeq \label{eq:M-inf-additive-network}
M_{\rm inf}=\frac{\bar{M}}{1-m\bar{K} \ln\left(1+\frac{1}{m\bar{K}}\right)}
\endeq
with $\bar{K}=\max_{i} K_i$ and $\bar{M}=\max_{i} M_i$.
\endth

\begpr
It is first shown that each generating series $d_{ji}$ is locally convergent. Consider the case where
every node series $c_i\in\re_{LC}\langle\langle X_i\rangle\rangle$ is a {\em maximal series}
$\bar{c}_i:=\sum_{\eta\in X^\ast}K_iM_i^{\abs{\eta}}\abs{\eta}!\, \eta$.
That is, every coefficient of $\bar{c}_i$ is growing at its maximal rate. While $y_i=F_{c_i}[u_i]$ may not have a finite
dimensional state space realization,
it is easily shown that a maximal series has the realization
\begdi 
\dot{z}_i=\frac{M_i}{K_i}z_i^2(1+u_i),\;\;z_i(0)=K_i,\;\;y_i=z_i
\enddi
\hspace*{-0.07in}\cite[Lemma 3]{Thitsa-Gray_12}.
Therefore, the corresponding network can be realized by
\begeq
\label{eq:maximal-series-state-space}
\dot{z}_i=\frac{M_i}{K_i}z_i^2\left(1+\sum_{j=1}^m W_{ij}z_j+v_i\right)\!\!,\;\;z_i(0)=K_i,\;\;y_i=z_i,
\endeq
$i=1,2,\ldots,m$. As this realization of the input-output map $v\mapsto y$ is polynomial, it is clearly real analytic.
Therefore, every generating series for $v_i\mapsto y_j$, say $\bar{d}_{ji}$, must be locally convergent \cite[Lemma 4.2]{Sussmann_83}.
The claim now is that $d_{ji}$ must also be locally convergent since $\abs{\langle d_{ji},\eta\rangle}\leq \langle \bar{d}_{ji},\eta\rangle$
for all $\eta\in X^\ast$. This inequality is most easily deduced from the formal realization of $v_i\mapsto y_j$ given in Theorem~\ref{th:additive-interconnections},
where the Lie derivatives used to compute the coefficients of $d_{ji}$ will all be upper bounded in magnitude by the Lie derivatives computed using maximal series.

Next, a suitable geometric growth constant for the network ${\mathcal N}_m$ is determined.
First observe that the growth constants $\bar{K}$ and $\bar{M}$ constitute a worst case maximum growth rate for every
node in the network.
In light of the formal representation of any $d_{ji}$ in Theorem~\ref{th:additive-interconnections}, the growth
rate of $d_{ji}$ is upper bounded by the growth rate of the natural response $\langle \bar{d}_{ji},x_0^k\rangle=L_{V_0}^k\hat{c}_j(\mbf{1})$, $k\geq 0$,
where $W_{ij}=1$ for all $i,j$, and every non-trivial component of $\hat{c}_j$ is the maximal series
$\bar{c}=\sum_{\eta\in X^\ast}\bar{K} \bar{M}^{\abs{\eta}}\abs{\eta}!\, \eta$.
(See \cite[Lemma 7]{Thitsa-Gray_12} for an alterative approach when $m=1,2$.)
From the symmetry of such a {\em maximal network}, $z_i=z_j$ for all $i,j$. Applying these conditions to \rref{eq:maximal-series-state-space},
the natural response at each node is
given by the solution of the Abel differential equation
\begeq \label{eq:maximal-network-Abel-equation}
\dot{z}=\frac{\bar{M}}{\bar{K}}(z^2+mz^3),\;\;z(0)=\bar{K}.
\endeq
It can be directly verified that this equation has the solution
\begdi
z(t)=\frac{-\frac{1}{m}}{1+\mathcal{W}\left[-\left(1+\frac{1}{m\bar{K}}\right)\exp\left(\frac{\bar{M}}{m\bar{K}}t-\left(1+\frac{1}{m\bar{K}}\right)\right)\right]},
\enddi
where $\mathcal{W}$ denotes the Lambert $W$-function, that is, the inverse of the function
$f(x)=x\exp(x)$ corresponding to the principal branch of this multi-valued function \cite{Corless-etal_96}.
As $\mathcal{W}$ is known to be holomorphic on the complex plane,
$z(t)$ will therefore be analytic at $t=0$. The corresponding Taylor series has a radius of convergence determined by the singularity nearest to the origin,
in this case
\begdi
t^\ast=\frac{1}{\bar{M}}\left(1-m\bar{K}\ln\left(1+\frac{1}{m\bar{K}}\right)\right).
\enddi
Applying a well known theorem from complex analysis (see \cite[Theorem~2.4.3]{Wilf_94}) gives
the infimum of all geometric growth constants for the maximal network, namely $M_{\rm inf}=1/t^\ast$.
(Note that the function $\lambda(x)=1-x\ln(1+{1/x})$ is a decreasing function, which further justifies using the maximum $K_i$ in
the network as the worst case.) Since for any $M>M_{\rm inf}$ there is a $K>0$ to upper bound the fastest coefficient growth
in the maximal network,
the generating series for every node in the original network must also be upper bounded by this growth rate.
\endpr

It is worth noting that \rref{eq:M-inf-additive-network} is in fact identical to the growth constant identified for
unity feedback systems with $m$ inputs as described in \cite[Corollary~2]{Thitsa-Gray_12}. While the network topologies
are clearly distinct, this point of tangency is derived from the fact that unity feedback systems and
additive maximal networks both have natural responses satisfying \rref{eq:maximal-network-Abel-equation}.

\begin{table}[tb]
\caption{Integer sequences generated by maximal additive network with unity growth constants}
\label{tbl:network-series}
\vspace*{-0.1in}
\begin{center}
\begin{tabular}{c|l|c|c}
	\toprule 
	$m$ & \hspace*{0.8in}$a_n$ & $M_{\rm inf}$ & $\hat{M}_n$ \\
	\midrule 
	1 & 1, 2, 10, 82, 938, 13778, 247210, $\ldots$            & 3.2589     & 3.22634 \\
	2 & 1, 3, 24, 318, 5892, 140304, $\ldots$         & 5.2891     & 5.23618 \\
	3 & 1, 4, 44, 804, 20556, 675588, $\ldots$        & 7.3017     & 7.22873 \\
	4 & 1, 5, 70, 1630, 53120, 2225480, $\ldots$      & 9.3088     & 9.21567 \\
	5 & 1, 6, 102, 2886, 114294, 5819190, $\ldots$    & 11.3132    & 11.2001 \\
	6 & 1, 7, 140, 4662, 217308, 13022688,$\ldots$   & 13.3163    &  13.1831 \\
	\bottomrule 
\end{tabular}
\end{center}
\end{table}

\begin{figure}[tb]
\begin{center}
\includegraphics[scale=0.55]{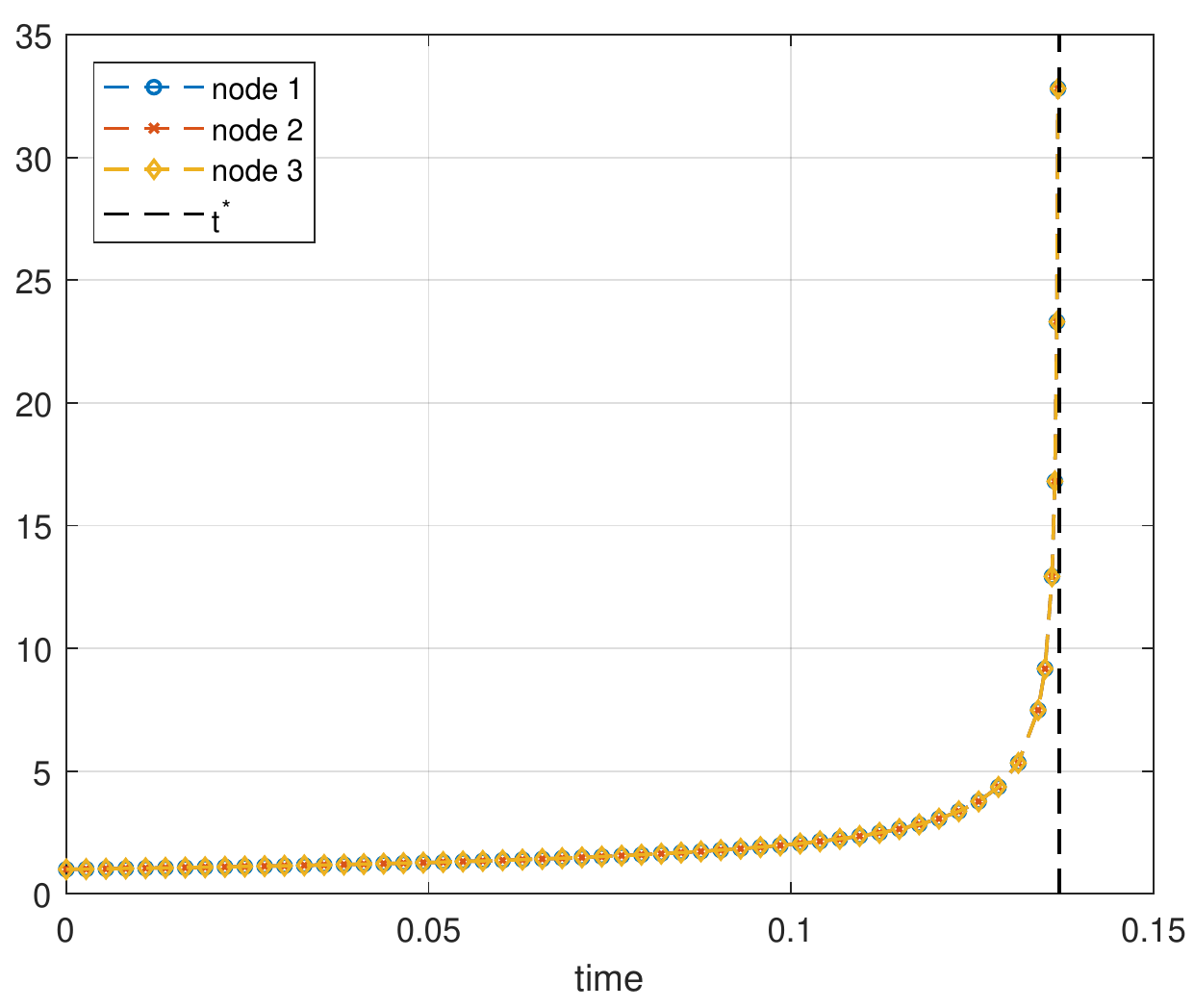}
\end{center}
\vspace*{-0.2in}
\caption{Natural response of three node maximal network in Example~\ref{ex:three-node-maximal-network-worst-case}.}
\label{fig:three-node-maximal-network-worst-case}
\end{figure}

\begex \label{ex:three-node-maximal-network-worst-case}
{\rm
Consider a maximal additive network where $K_i=M_i=1$, $i=1,2,\ldots,m$. The Taylor series
of the natural response has integer coefficients $a_n$, $n\geq 1$ as shown in Table~\ref{tbl:network-series}.
The coefficients when $m=1$ correspond to the OEIS integer sequence
A112487 \cite{OEIS}. The table also shows the growth rate $M_{\rm \inf}$ computed from \rref{eq:M-inf-additive-network}
and an estimate of the growth constant $M$ computed
from $\hat{M}_n=na_{n}/a_{n-1}$ when $n=50$.
The corresponding three node network was simulated in MatLab for the zero input case. The node responses,
which are identical, are
shown in Figure~\ref{fig:three-node-maximal-network-worst-case}.
Since the coefficients of every generating series are positive,
it is known that the natural response of every node will have
a finite escape time at $t=t^\ast$ (see \cite[Theorem 11]{Thitsa-Gray_12}).
In this case, $t^\ast=1/M_{\rm inf}=0.1379$, which is what was observed
in the simulation.
}
\endex

\begin{figure}[ht]
\begin{center}
\includegraphics[scale=0.55]{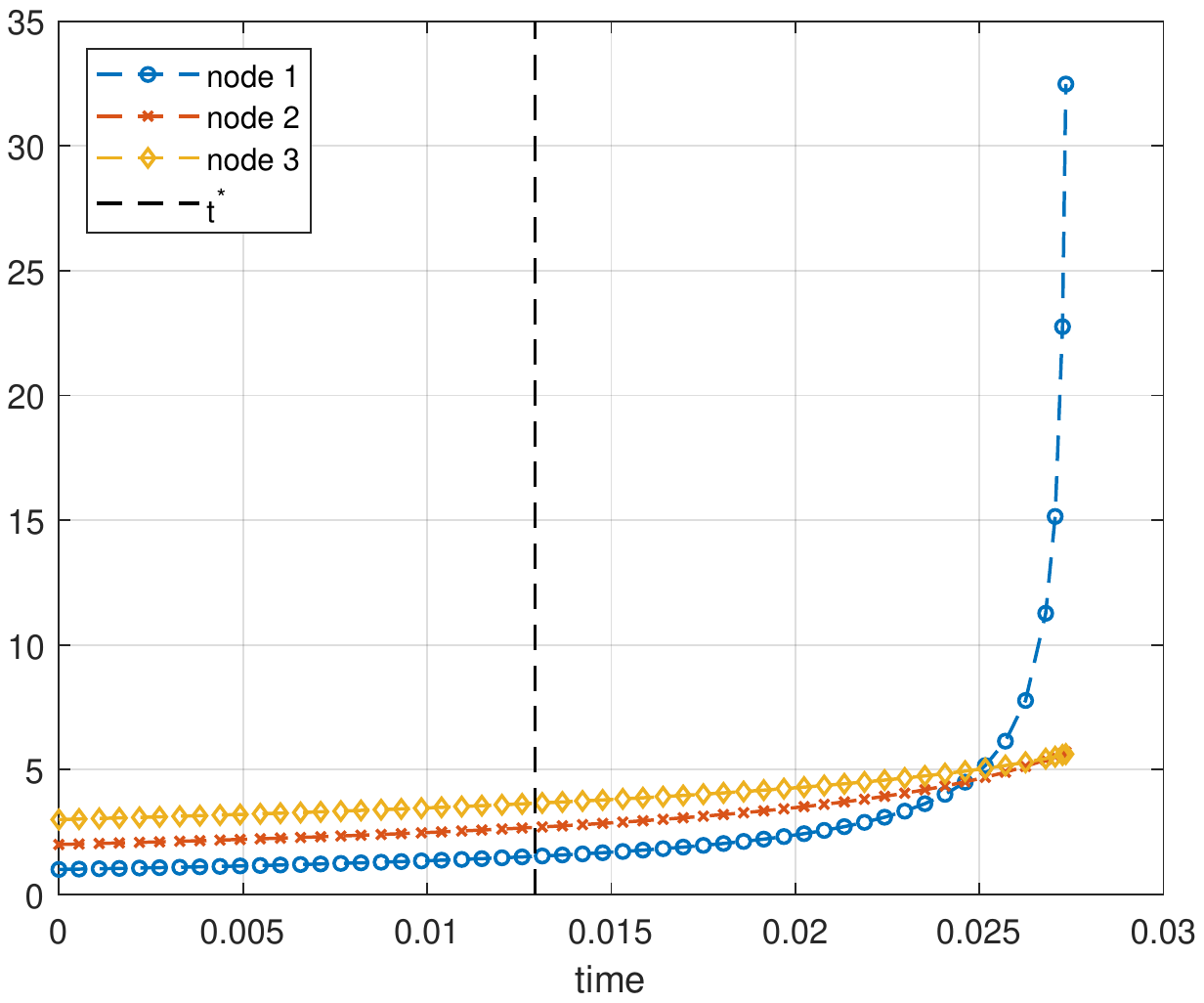}
\end{center}
\vspace*{-0.1in}
\caption{Natural response of three node network in Example~\ref{ex:three-node-maximal-network-generic}.}
\label{fig:three-node-maximal-network-generic}
\end{figure}

\begex \label{ex:three-node-maximal-network-generic}
{\rm
Consider a three node additive network involving maximal series with $K_i=i$, $M_i=5-i$ and
\begdi
W=\left[\begin{array}{cccc}
1 & 0.5 & 1  \\
1 & 1 & 0 \\
0.25 & 1 & 1
\end{array}\right].
\enddi
Thus, $\bar{K}=3$, $\bar{M}=4$, and $M_{\rm inf}=77.2867$.
The node natural responses are
shown in Figure~\ref{fig:three-node-maximal-network-generic}.
As this network is not maximal,
$t^\ast=1/M_{\rm inf}=0.01294$ provides only a lower bound on the escape times
of each node.
}
\endex

\section{Additive Networks of Chen-Fliess Series: Relative Degree}
\label{sec:relative-degree}

In this section the following question is addressed: When does the generating series of the
mapping $v_i\mapsto y_j$ in an additively interconnected
network ${\mathcal N}_m$ have a well defined relative degree?
The treatment starts with the easiest case first as described next.
It is assumed throughout that ${\mathcal N}_m$ is
comprised of systems with generating series $c_i$ which have relative degree $r_i$ for
$i=1,2,\ldots,m$.

\begde
The $i$-th node in a network ${\cal N}_m$ is said to be {\em fully connected} if $W_{ij}\neq 0$ for all $j\neq i$.
A network ${\cal N}_m$ is said to be {\em fully connected} if every node is fully connected.
\endde

Note that self-loops, i.e., when $W_{ii}\neq 0$, are not important in the present context as proportional output feedback
is easily shown to preserve relative degree \cite{Gray-Venkatesh_SCL19}.

\begth \label{th:relative-degree-fully-connected}
If
the $i$-th node in ${\mathcal N}_m$ is fully connected, then the generating series $d_{ji}$ for mapping
$v_i\mapsto y_j$ has relative degree $r_{ji}=r_j+r_i$.
\endth

\begpr
Observe that the full output at node $j$ is
\begin{align*}
y_j&=F_{c_j}\left[v_j+\sum_{k=1}^m W_{jk} y_k \right] \\
&=F_{c_j}\left[v_j+\sum_{k,l=1}^m W_{jk} F_{d_{kl}}[v_l]\right].
\end{align*}
For any $i\neq j$, that part of $y_j$ in response to $v_i$ acting alone (i.e., $v_l=0$ for $l\neq i$) is
given by
\begin{align*}
y_{j}&=F_{c_j}\Bigg[W_{ji}F_{c_i}[v_i]+\sum_{k=1\atop k\neq i}^m W_{jk} F_{d_{ki}}[v_i]+ \\
&\hspace{0.2in} \sum_{k,l=1\atop l\neq i}^m W_{jk} F_{d_{kl}}[0]\Bigg].
\end{align*}
Note that for all $k\neq i$, $\supp(d_{ki})\subseteq x_0^{r}X^\ast$, where $r\geq r_i+1$, since
$v_i$ passes through $F_{c_i}$ in every path leading to the $j$-th node. In which case, the argument of $F_{c_j}$ above
has a generating series with relative degree $r_i$. The conclusion then follows immediately from Theorem~\ref{th:relative-degree-casecade}.
\endpr

\begin{figure}[ht]
\begin{center}
\includegraphics[scale=0.7]{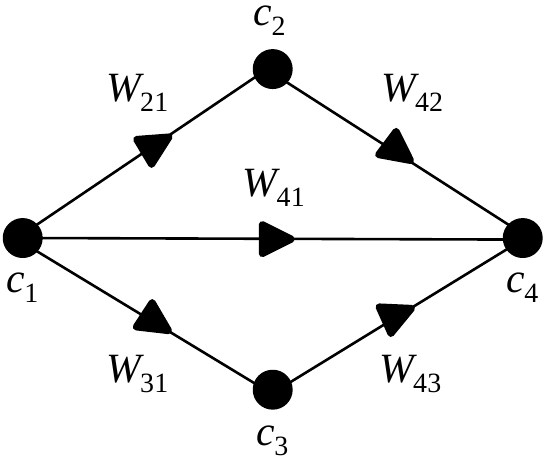}
\end{center}
\vspace*{-0.2in}
\caption{Four node network in Example~\ref{ex:four-node-counterexample}.}
\label{fig:four-node-counterexample}
\end{figure}

\begex \label{ex:four-node-counterexample}
{\rm
Consider the network shown in Figure~\ref{fig:four-node-counterexample}.
The corresponding weighting matrix is
\begdi
W=\left[\begin{array}{cccc}
0 & 0 & 0 & 0 \\
W_{21} & 0 & 0 & 0 \\
W_{31} & 0 & 0 & 0 \\
W_{41} & W_{42} & W_{43} & 0
\end{array}\right].
\enddi
The network is clearly {\em not} fully connected,
but node $4$ is fully connected assuming $W_{4j}\neq 0$, $j=1,2,3$.
Therefore, applying the theorem above gives, for example, that
$r_{41}=r_4+r_1$.

Suppose now that $W_{41}=0$ so that the theorem no longer applies. Further assume that $r_2=r_3=r$.
Observe that
\begdi
u_4=W_{42}F_{c_2}[W_{21}F_{c_1}[u_1]]+W_{43}F_{c_3}[W_{31}F_{c_1}[v_1]],
\enddi
and thus,
\begdi
d_{41}=c_4\circ [W_{42}(c_2\circ (W_{21}c_1))+W_{43}(c_3\circ (W_{31}c_1))].
\enddi
Both $c_2\circ (W_{21}c_1)$ and $c_3\circ (W_{31}c_1)$ have relative degree $r+r_1$,
but $d_{41}$ can fail to have relative degree.
As a simple example, suppose $c_1=c_2=c_4=x_1$ and $c_3=-x_1$ so that
$d_{41}=(W_{42}W_{21}-W_{43}W_{31}) x_0^2x_1$.
If $W$ is such that $W_{42}W_{21}=W_{43}W_{31}$, then $d_{41}=0$ does not have relative degree.
On the other hand, if the symmetry condition $r_2=r_3$ is broken, then it follows that $d_{41}$ has
relative degree $r_{41}=r_4+\min (r_2,r_3)+r_1$.
}
\endex

The final case in the example above suggests a sufficient condition for the general case. Namely,
in the absence of these degenerate situations where a node is presented with an input whose underlying
generating series does not have relative degree, the relative degree for $d_{ji}$ will be well defined and determined
by a path from node $i$ to node $j$ whose {\em accumulated} relative degrees is minimal. To make this
claim more precise, the following language adapted from signal flow graph theory will be useful.

Let ${\mathcal N}_m$ be a given additive network.
An {\em edge} is a directed line segment connecting two nodes.
A {\em path} is a continuous set of edges connecting two nodes in ${\mathcal N}_m$ and traversed in the direction indicated.
A {\em forward path} is a path in which no node is encountered more than once.
A {\em loop} is a path that originates and ends on the same node in which no node is encountered more than once.
Finally, the {\em subgraph} $G_{ji}$ from node $i$ to node $j$ is the simple graph (i.e., all loops are omitted)
consisting of all forward paths connecting node $i$ and node $j$.

The following theorems provide a sufficient condition under which the relative degree is well defined for a given
input-output map $v_i\mapsto y_j$ in an additive network. Given a subgraph $G_{ji}$, the {\em accumulated relative degree} of node $i$ is
$r_i^+=r_i$. If node $k\neq i$ in $G_{ji}$ has $N$ incoming edges from nodes $i_1,i_2,\ldots, i_N$ with accumulated relative degrees
$r^+_{i_1},r^+_{i_2},\ldots,r^+_{i_N}$, respectively, then the {\em accumulated relative degree} at node $k$ is
\begdi
r^+_k=r_k+\min\{r^+_{i_1},r^+_{i_2},\ldots,r^+_{i_N}\}.
\enddi
Note this definition does not imply that any mappings defined by the network have relative degree, it simply
computes the {\em potential} relative degree of such a mapping should it be well defined.

\begth \label{th:relative-degree-additive-network-distinct-condition}
Let $i$ and $j$ be fixed nodes in ${\mathcal N}_m$.
If at every node $l\notin\{i,j\}$  the accumulated relative degrees
of the nodes from every incoming edge are distinct, then
the generating series $d_{ji}$ for $v_i\mapsto y_j$ in ${\mathcal N}_m$ has well defined relative degree
equivalent to $r_{ji}=r_j^+$.
\endth

\begpr
As feedback loops do not affect the relative degree of any forward path, it is sufficient
to consider only the subgraph $G_{ji}$.
The claim then follows directly from Corollary~\ref{co:relative-degree-multi-sum-distinct},
Theorem~\ref{th:relative-degree-casecade}, and the definition of accumulated relative degree.

\endpr

The distinctness condition in the above theorem can be relaxed by
utilizing instead the condition in Corollary~\ref{co:relative-degree-multi-sum-not-distinct}.

\begth \label{th:relative-degree-additive-network-repeated-condition}
Let $i$ and $j$ be fixed nodes in ${\mathcal N}_m$.
If at every node $l\notin\{i,j\}$ the accumulated relative degrees
of the nodes from every incoming edge satisfy the condition in Corollary~\ref{co:relative-degree-multi-sum-not-distinct},
then the generating series $d_{ji}$ for $v_i\mapsto y_j$ in ${\mathcal N}_m$ has well defined relative degree
equivalent to $r_{ji}=r_j^+$.
\endth

\begin{figure}[ht]
\begin{center}
\includegraphics[scale=0.7]{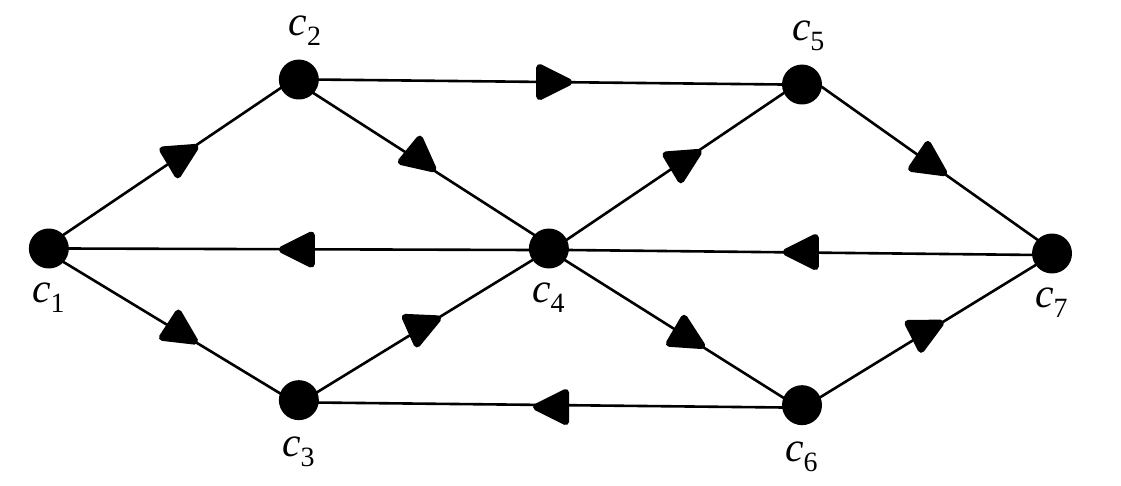}
\end{center}
\vspace*{-0.1in}
\caption{Network in Example~\ref{ex:double-diamond-example}.}
\label{fig:double-diamond-example}
\end{figure}

\begin{figure}[ht]
\begin{center}
\includegraphics[scale=0.7]{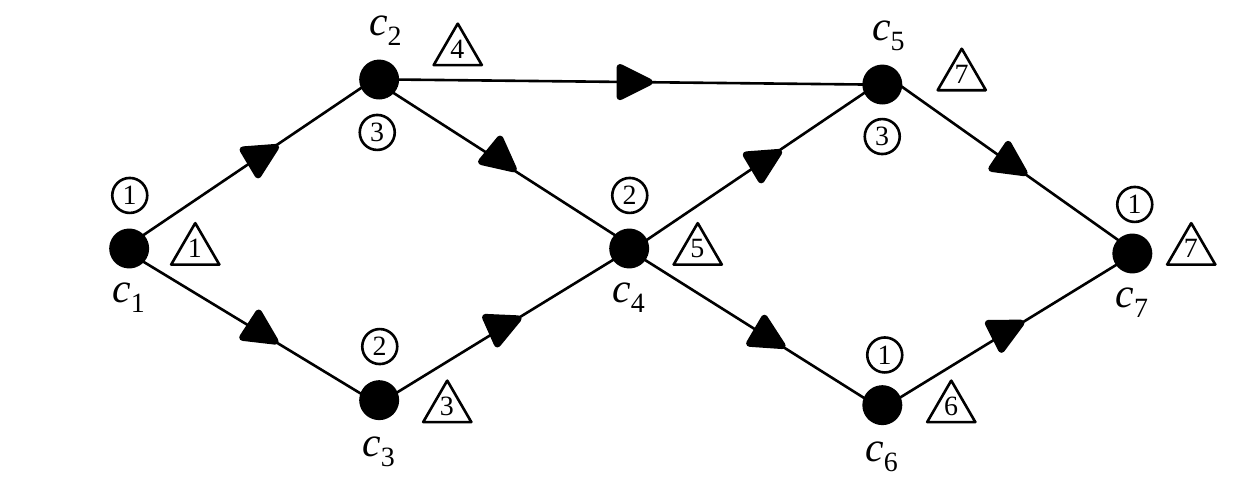}
\end{center}
\vspace*{-0.1in}
\caption{Subgraph of forward paths for $v_1\mapsto y_7$ in Example~\ref{ex:double-diamond-example}. The relative
degree of each generating series $c_i$ is the circled number. The accumulated relative degree at each node is the number in the triangle.}
\label{fig:double-diamond-example-subgraph}
\end{figure}

\begex \label{ex:double-diamond-example}
{\rm
Consider the network shown in Figure~\ref{fig:double-diamond-example}, where each weight $W_{ij}\in \{0,1\}$ (i.e., $0\sim\mbox{not connected}$,
$1\sim\mbox{connected}$), and
the generating series for the nodes are:
\begin{align*}
c_1 &= K_1 x_1 + 2 x_0  x_1 \\
c_2 &= x_0+K_2 x_0^2  x_2 \\
c_3 &= K_3 x_0  x_3 + 3 x_0^2  x_3^2 \\
c_4 &= 1+K_4 x_0  x_4 - x_0^2  x_4x_0 \\
c_5 &= 4x_0+K_5 x_0^2  x_5 -2x_0^4x_5 \\
c_6 &= K_6 x_6 -x_6^2\\
c_7 &= x_0+2+K_7 x_7+4x_0x_7
\end{align*}
with $K_i\neq 0$ in every case.
The subgraph of forward paths is shown in Figure~\ref{fig:double-diamond-example-subgraph}.
The relative degree of the generating series at each node is the circled number shown next to each node.
The accumulated relative degree at each node is the number in the triangle
The goal is to determine the relative degree of the mapping $v_1\mapsto y_7$, provided it is well defined.
Observe that only nodes $4$, $5$ and $7$, have more than one incoming edge. In each case, the accumulated relative degrees
are distinct, namely, $3,4$; $4,5$; and $6,7$, respectively. Therefore, Theorem~\ref{th:relative-degree-additive-network-distinct-condition}
applies,
and $r_{71}=7$.
To independently verify this claim,
the generating series $d_{71}$ was computed using
the full network via Theorem~\ref{th:additive-interconnections} with the aid of Mathematica and found to be
\begdi
d_{71}=d_{71,N}+K_1K_3K_4K_6K_7x_0^6x_1+x_0^6e,
\enddi
where
\begin{align*}
d_{71,N} = & {\,} x_0+
(4 K_7 + K_6 K_7)x_0^2  +(16 + 4 K_6 - K_7)x_0^3+ \\
& {\,} (-4 + K_5 K_7 + 2 K_4 K_6 K_7) x_0^4 + (4 K_5 + 8 K_4 K_6- \\
& {\,} 8 K_4 K_7 + K_5 K_7 + 2 K_4 K_6 K_7) x_0^5 + \cdots\\
e = & {\,} (4 K_1 K_3 K_4 K_6 - 6 K_1 K_3 K_4 K_7 + K_1 K_2 K_5 K_7+ \\
& {\,}K_1 K_2 K_4 K_6 K_7 + 2 K_3 K_4 K_6 K_7) x_0 x_1 + \cdots
\end{align*}
The relative degree of $d_{71}$ is 7 as expected.
}
\endex

An additive network ${\mathcal N}_m$ is said to have {\em complete relative degree} if every mapping
$v_i\mapsto y_j$, $i,j=1,2,\ldots,m$ has relative degree. From Theorem~\ref{th:relative-degree-fully-connected} it
is immediate that fully connected networks have this property. Another class of networks sharing this property is given
in the following theorem. It states that in some sense the property of a network having complete relative degree is {\em generic}.

\begth
Consider an additive network ${\mathcal N}_m$ where the weighting matrix has entries $W_{ij}\in\{0,1\}$. If the
unity weights are replaced with continuous random variables, then every sample network has complete relative degree.
\endth

\begpr
At any given node, the incoming nodes may or may not have distinct accumulated relative degree. In the case where they do,
then Theorem~\ref{th:relative-degree-additive-network-distinct-condition} applies, otherwise, Theorem~\ref{th:relative-degree-additive-network-repeated-condition}
applies provided the condition for multiplicities greater than one can be met. Specifically, at node $k$ with incoming edges from nodes
$j_1,j_2,\ldots, j_N$ with accumulated relative degrees
$r^+_{j_1},r^+_{j_2},\ldots,r^+_{j_N}$, it is required that if $r^+_j$ is repeated $s_j>1$ times then
\begin{align*}
&W_{ij(1)}\langle d_{j(1)i},x_0^{r^+_j-1}x_i\rangle+
W_{ij(2)} \langle d_{j(2)i},x_0^{r^+_j-1}x_i\rangle+
\cdots \\
&\hspace*{0.2in}+W_{ij(s_j)}\langle d_{j(s_j)i},x_0^{r^+_j-1}x_i\rangle\neq 0,
\end{align*}
where $j(l)\in\{j_1,j_2,\ldots, j_N\}$, and the $W_{ij_l}$ are random variables with any continuous distribution(s).
But this condition is always true with probability one, and hence, the theorem is proved.
\endpr

\begin{figure}[ht]
\begin{center}
\includegraphics[scale=0.6]{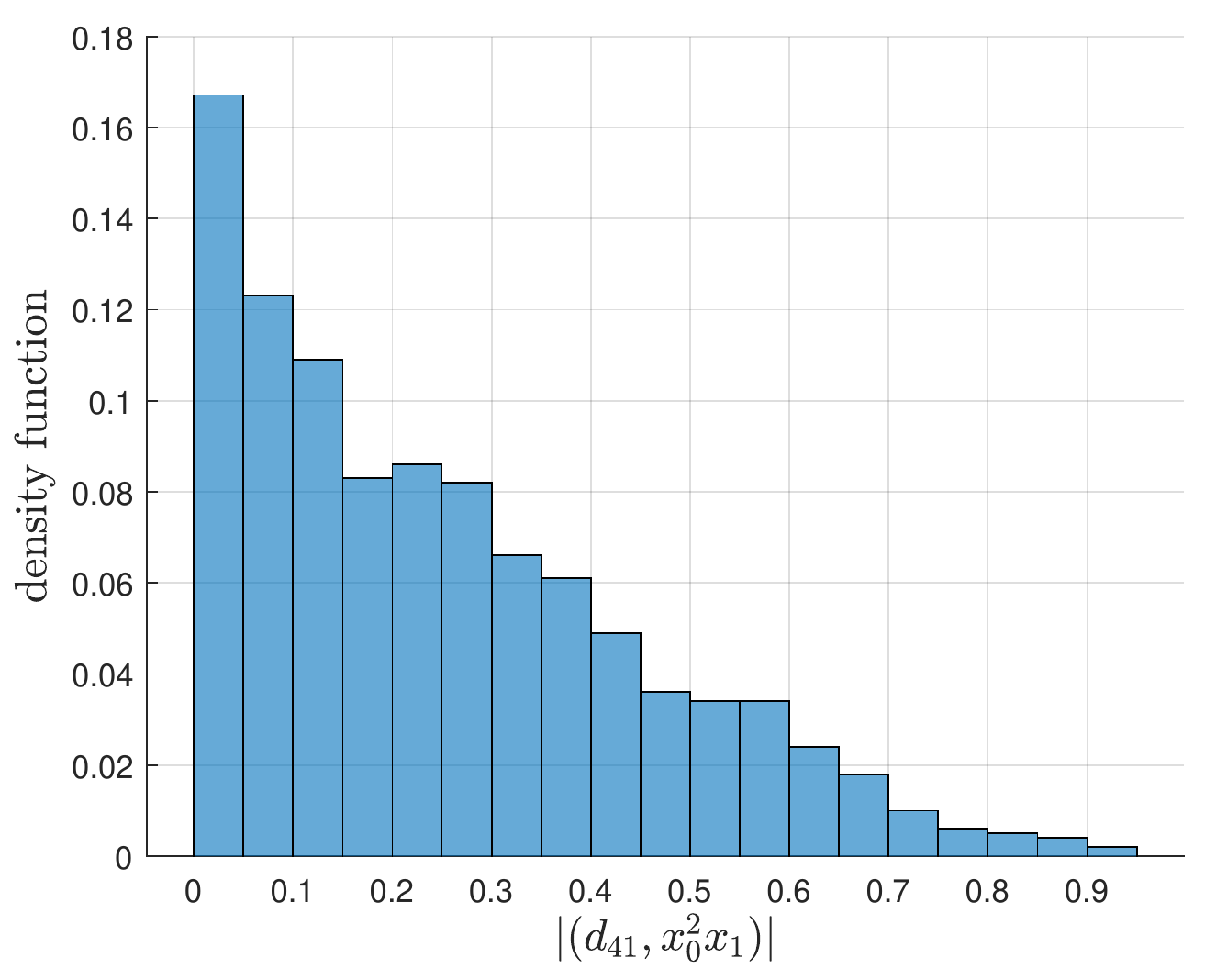}
\end{center}
\vspace*{-0.1in}
\caption{Estimate of density function for $\abs{\langle d_{41},x_0^2x_1\rangle}$ in Example~\ref{exer:random-network}.}
\label{fig:random-network-density-function}
\end{figure}

\begex \label{exer:random-network}
{\rm
Reconsider Example~\ref{ex:four-node-counterexample}, where $W_{41} = 0$
and now $W_{21}$, $W_{31}$, $W_{42}$ and $W_{43}$ are i.i.d.
random variables with a uniform distribution on $[0,1]$.
An estimate of the density function for the random variable $\abs{\langle d_{41},x_0^2x_1\rangle}$ is shown in
Figure~\ref{fig:random-network-density-function}. In every case of the 1000
random networks generated, $d_{41}$ had relative degree $r=3$ as expected.
}
\endex

\section{Conclusions}

Two basic properties were established for an additive network of input-output systems where each
node of the network is modeled by a convergent Chen-Fliess series.
First it was shown that every input-output map between a pair of nodes has
a locally convergence Chen-Fliess series representation. An explicit and in some cases achievable growth bound on the
coefficients was computed using the notion of a maximal network. Second, sufficient
conditions were given under which the input-output map between a pair
of nodes has a well defined relative degree as defined by
its generating series. This analysis led to the conclusion
that this relative degree property is generic when the connection
strengths between nodes are randomized.

%


\begin{thebibliography}{99}


%

\bibitem{Corless-etal_96}
R. M. Corless, G. H. Gonnet, D. E. G. Hare, D. J. Jeffrey, and D. E. Knuth,
On the Lambert $W$ function, {\em Adv. Comput. Math.}, 5 (1996) 329--359.

\bibitem{Ferfera_80}
A.~Ferfera, Combinatoire du mono\"{i}de libre et composition de
certains syst\`{e}mes non lin\'{e}aires, {\em Ast\'{e}risque},
75-76 (1980) 87--93.

\bibitem{Fliess_81}
M.~Fliess, Fonctionnelles causales non lin\'{e}aires et ind\'{e}termin\'{e}es non
commutatives, {\em Bull.~Soc.~Math.~France}, 109 (1981) 3--40.

\bibitem{Fliess_83}
M.~Fliess, R\'{e}alisation locale des syst\`{e}mes non lin\'{e}aires,
alg\`{e}bres de Lie filtr\'{e}es transitives et s\'{e}ries
g\'{e}n\'{e}ratrices non commutatives, {\em Invent.~Math.}, 71 (1983) 521--537.

\bibitem{Golubitsky-Stewart_06}
M. Golubitsky and I. Stewart,
Nonlinear dynamics of networks: The groupoid formalism,
{\em Bull. Amer. Math. Soc. (N.S.)}, 43 (2006) 305--364.


\bibitem{Gray-etal_AUTO14}
W.~S.~Gray, L.~A.~Duffaut Espinosa, and M.~Thitsa,
Left inversion of analytic nonlinear SISO systems via formal power series
methods, {\em Automatica}, 50 (2014) 2381--2388.

\bibitem{Gray-Ebrahimi-Fard_SIAM17}
W.~S.~Gray and K.~Ebrahimi-Fard,
SISO output affine feedback transformation group and its Fa\`{a} di Bruno Hopf algebra,
{\em SIAM J.~Control Optim.}, 55 (2017) 885--912.

\bibitem{Gray-Ebrahimi-Fard_SCL21}
W. S. Gray and K. Ebrahimi-Fard,
Generating series for networks of Chen-Fliess series,
{\em Systems Control Lett.}, 147 (2021) article 104827.

\bibitem{Gray-Venkatesh_SCL19}
W. S. Gray and G. S. Venkatesh,
Relative degree of interconnected SISO nonlinear control systems,
{\em Systems Control Lett.}, 124 (2019) 99--105.

\bibitem{Gray-etal_Auto20}
W.~S.~Gray, G.~S.~Venkatesh, and L.~A.~Duffaut Espinosa,
Nonlinear system identification for multivariable
control via discrete-time Chen--Fliess series,
{\em Automatica}, 119 (2020) 109085.

\bibitem{Gray-Wang_02}
W.~S.~Gray and Y.~Wang,
Fliess operators on $L_p$ spaces: Convergence and continuity,
{\em Systems Control Lett.}, 46 (2002) 67--74.



\bibitem{Isidori_95}
A.~Isidori, {\em Nonlinear Control Systems}, 3rd Ed., Springer, London, 1995.

\bibitem{Jiang-Lai_19}
J. Jiang and Y.-C. Lai,
Irrelevance of linear controllability to nonlinear dynamical networks,
{\em Nature Communications}, 2019, https://doi.org/10.1038/s41467-019-11822-5.

\bibitem{Kawski-Sussmann_97}
M.~Kawski and H.~J.~Sussmann,
Noncommutative power series and formal {L}ie-algebraic techniques in nonlinear control theory,
in {\em Operators, Systems, and Linear Algebra: Three Decades of Algebraic Systems Theory},
U.~Helmke, D.~Pr{\"a}tzel-Wolters, and E.~Zerz, Eds.,
B.~G.~Teubner, Stuttgart, 1997, pp.~111--128.


\bibitem{Menara-etal_CDC20}	
T. Menara, G. Baggio, D. S. Bassett, and F. Pasqualetti,
Conditions for feedback linearization of network systems,
{\em IEEE Control Systems Letters}, 4 (2020) 578--583.

\bibitem{Nijmeijer-vanderSchaft_90}
H.~Nijmeijer and A.~J.~van~der Schaft,
{\em Nonlinear Dynamical Control Systems},
Springer, New York, 1990.

\bibitem{OEIS}
N.~J.~A.~Sloane, {\em The On-Line Encyclopedia of Integer Sequences}, available
at https://oeis.org.



\bibitem{Sussmann_83}
H.~J.~Sussmann,
Lie brackets and local controllability: A sufficient condition for scalar-input systems,
{\em SIAM J. Control Optim.}, 21 (1983) 686--713.


\bibitem{Thitsa-Gray_12}
M.~Thitsa and W.~S.~Gray,
On the radius of convergence of interconnected analytic nonlinear input-output systems,
{\em SIAM J. Control Optim.}, 50 (2012) 2786--2813.

%


\bibitem{Wilf_94}
H. S. Wilf, {\em Generatingfunctionology}, 2nd Ed., Academic Press, San Diego, CA, 1994.

\end{thebibliography}
\end{document}